\documentclass{aa}  

\usepackage{graphicx}
\usepackage[varg]{txfonts}
\usepackage[normalem]{ulem}
\usepackage[usenames]{color}
\usepackage{mathrsfs}
\usepackage{natbib}
\usepackage{enumitem}

\usepackage{geometry}                           % See geometry.pdf to learn the layout options. There are lots.
\usepackage{graphicx}                           % Use pdf, png, jpg, or eps§ with pdflatex; use eps in DVI mode
\usepackage{amssymb}
\usepackage{nccmath}
\usepackage{booktabs}
\usepackage[tableposition=top]{caption} % Spaces the caption properly
\usepackage{multirow}
\usepackage{hyperref}
\usepackage[font=small,labelfont=bf]{caption}
\usepackage{media9}
\newcommand{\Mm}{{\mathrm{\, Mm}}}
\newcommand{\kms}{{\mathrm{\, km \,s^{-1}}}}

\newcommand{\secs}{{\mathrm{\, seconds}}}
\newcommand{\mins}{{\mathrm{\, minutes}}}

\definecolor{valeriia}{rgb}{1.0, 0.0, 0.0}
\definecolor{manuelcomment}{rgb}{0.93, 0.57, 0.13}
\definecolor{manuel}{rgb}{0.3, 0.3, 0.9}
\definecolor{elena}{rgb}{0.0,0.5,0.0}
\newcommand{\notesize}{\fontsize{4}{10}\selectfont} %<-- New fontsize
\def\overbigdot#1{\overset{\hbox{\notesize$\bullet$}}{#1}}

\begin{document} 

\title{Large-amplitude longitudinal oscillations in solar prominences simulated with different resolutions}

   \author{V. Liakh\inst{1, 2}
   \and 
   M. Luna \inst{3, 4}
   \and 
   E. Khomenko\inst{1, 2}}

 \institute{Instituto de Astrof\'{\i}sica de Canarias, E-38205 La Laguna, Tenerife, Spain\\
              \email{vliakh@iac.es}
         \and
             Departamento de Astrof\'{\i}sica, Universidad de La Laguna, E-38206 La Laguna, Tenerife, Spain\\
         \and
             Departament de F\'{\i}sica, Universitat de les Illes Balears, E-07122, Palma de Mallorca, Spain\\
         \and
             Institute of Applied Computing \& Community Code (IAC$^3$), UIB, Spain
             }
   
%   \institute{Instituto de Astrof\'{\i}sica de Canarias, E-38200 La Laguna, Tenerife, Spain  \label{inst1}\centering
%   \and Universidad de La Laguna, Dept. Astrof\'{\i}sica, E-38206 La Laguna, Tenerife, Spain\\  \email{vliakh@iac.es} \label{inst2}\centering}

 \date{}

 \titlerunning{Numerical simulations of LAOs}
\authorrunning{Liakh, Luna \& Khomenko}

\abstract
  {Large-amplitude longitudinal oscillations (LALOs) in solar prominences have been widely studied in the last decades. However, their damping and amplification mechanisms are not well understood.}
  {In this study, we investigate the attenuation and amplification of LALOs using high-resolution numerical simulations with progressively increasing spatial resolutions.}
  {We performed time-dependent numerical simulations of LALOs using the 2D magnetic configuration that contains a dipped region. After the prominence mass loading in the magnetic dips, we triggered LALOs by perturbing the prominence mass along the magnetic field. We performed the experiments with four values of spatial resolution.}
  {In the simulations with the highest resolution, the period shows a good agreement with the pendulum model. The convergence experiment revealed that the damping time saturates at the bottom prominence region with improving the resolution, indicating the existence of a physical reason for the damping of oscillations. At the prominence top, the oscillations are amplified during the first minutes and then are slowly attenuated. The characteristic time suggests more significant amplification in the experiments with the highest spatial resolution. The analysis revealed that the energy exchange between the bottom and top prominence regions is responsible for the attenuation and amplification of LALOs.}
  {The high-resolution experiments are crucial for the study of the periods and the damping mechanism of LALOs. The period agrees with the pendulum model only when using high enough spatial resolution. The results suggest that numerical diffusion in simulations with insufficient spatial resolution can hide important physical mechanisms, such as amplification of oscillations.}

% 5 {} token are mandatory
 
   \maketitle
%
%-------------------------------------------------------------------
 %
%-------------------------------------------------------------------

\section{Introduction}\label{sec:introduction}

Solar prominences are clouds of cold and dense plasma suspended in the solar corona, supported against gravity by the magnetic field. They are subject to various types of oscillations. These periodic motions are classified as small-amplitude and large-amplitude oscillations according to their velocity \citep[see reviews by][]{Oliver:2002solphys,Arregui:2018spr}. In the large-amplitude oscillations (LAOs), a large portion of the prominence oscillates with amplitudes that exceed $10\kms$. If the plasma motions are directed along the magnetic field, the oscillations are classified as large-amplitude longitudinal oscillations (LALOs). On the contrary,  oscillations with velocity directed perpendicular to the magnetic field are called transverse LAOs. More on the classification and properties of the different types of oscillations can be found in the last update of the living review by \citet{Arregui:2018spr}. 

\citet{Jing:2003apjl, Jing:2006solphys} have been first reported the LALOs in solar prominences. Since then, many observations provided important information on the properties of LALOs \citep[see, e.g.,][]{Vrsnak:2007aap, Zhang:2012aap, Li:2012apj, Luna:2014apj, Bi:2014apj, Zhang:2017apj}. Observations have shown that the period of the LALOs is around $1$ hour, and the damping time is usually less than two periods, indicating strong attenuation of the motions. \citet{Luna:2018apjs} cataloged the filaments oscillations using several months of the Global Oscillation Network Group (GONG) H$\alpha$ data near the maximum of the solar cycle $24$. The catalog provided a large sample of the LALOs and statistics of the periods and damping times. It was obtained that, on average, the ratio of damping time to period is equal to $1.25$, in agreement with the earlier observations.
An intriguing phenomenon in the prominence oscillations is the amplification of motions. Contrary to the damping caused by energy loss, oscillations somehow acquire additional energy and increase their amplitudes in time. The first amplification event was reported by \citet{Molowny-Horas:1999JOSO} from time series of H$\beta$ filtergrams of a polar crown prominence. They measured periodic Doppler signal and found clear oscillations with periods between 60 to 95 minutes and damping times between $119$ to $377$ minutes at six locations of the prominence. At one of the locations, they found an oscillation with a period of $28$ minutes and a growing amplitude with a characteristic time of $140$ minutes. More recently, \citet{Zhang:2017apj} reported observation of LALOs, in which the oscillatory amplitude remains constant or increases with time in some regions of the prominence. This unusual behavior of the LALOs amplitude has also been found in the GONG catalog by \citeauthor{Luna:2018apjs} (2018; see, e.g., Fig. 25). 

In the last decades, the LALOs have been studied in many analytical and numerical works. These were dedicated to answering questions related to the restoring and damping mechanisms. \citet{Luna:2012apjl} proposed a so-called pendulum model where the main contribution to the restoring force is the gravity projected along the magnetic field. This model suggests that the period is defined by the radius of curvature of the magnetic field \citep[see also][for another description of the model]{roberts_mhd_2019}. The following studies, based on the analytical calculations or the 2D and 3D simulations, confirmed that the pendulum model describes quite well the periods of longitudinal oscillations under typical prominence conditions \citep[e.g.,][]{Luna:2012apj, Luna:2016apj, Zhou:2018apj, Zhang:2019apj, Liakh:2020aap, Fan:2020apj}. \citet{Zhou:2018apj} found a systematic discrepancy of some 10\% with respect to the pendulum model. The authors suggested that the longitudinal oscillations produce dynamical deformations of the field structure. Under these circumstances, the period of LALOs can be slightly longer than predicted by the pendulum model due to the flattening of the magnetic field by the plasma motions. \citet{Zhang:2019apj} found that in the regime of a relatively weak magnetic field, when the gravity to Lorentz force ratio is close to unity, the trajectory of the prominence plasma is not along the rigid field lines. The authors found that these dynamic deformations of the magnetic structure contribute to the damping of the oscillations. Similarly, \citet{Fan:2020apj} found discrepancies with respect to the pendulum model also associated with the dynamical deformation of the field lines during the oscillations.

There have been proposed many candidates to damping mechanisms in order to explain the observed LALOs attenuation. Radiative losses and thermal conduction have been considered by \citet{Zhang:2012aap} and \citet{Zhang:2019apj} as damping mechanisms for LALOs. These authors pointed out the importance of the inclusion of the nonadiabatic effects into consideration. However, using numerical simulations, these authors have shown that nonadiabatic effects alone are not enough to explain the observed damping \citep[see also][]{Zhang:2020aap}.
\citet{Zhang:2013aap} and \citet{Ruderman:2016aap} proposed the alternative mechanism for the attenuation of LALOs. \citet{Zhang:2013aap} considered a possibility of LALOs decaying due to the mass drainage, while \citet{Ruderman:2016aap} studied the damping due to the mass accretion. Both studies showed that either the mass drainage or mass accretion results in decreasing velocity of the prominence. As a consequence, this leads to the damping of oscillations. \citet{Zhang:2019apj} found that in a situation of a relatively weak magnetic field, the wave leakage is an important mechanism for LALOs damping, in addition to the nonadiabatic effects. The motion of the prominence mass produces perturbations in the field structure that emanate in the form of MHD waves. 

In order to explain both the damping and the amplification of prominence oscillations, \citet{Ballester:2016aap} derived an expression for the temporal variation of the background temperature, taking into account the radiative losses and thermal conduction. The authors concluded that using some combination of the characteristic times of the different mechanisms and increasing or decreasing the background temperature could lead to the damping or amplification of the oscillations. \citet{Zhang:2017apj} proposed an alternative mechanism to explain the amplification: the threads located at the different dips of the same flux tube interchange their energy. Using 1D numerical simulations, \citet{Zhou:2017apj} studied the different combinations of the active-passive threads and demonstrated that the energy exchange significantly affects the damping or amplification time of the LALOs. However, this mechanism cannot explain the amplification of the threads that belong to the different field lines.

All the studies of LALOs in 2D and 3D have been done using numerical simulations. In these numerical experiments, dissipation is unavoidable.
\citet{Terradas:2016apj} investigated the influence of the numerical dissipation on the prominence oscillations in the 3D model, increasing spatial resolution up to $300$ km. They concluded that the energy losses associated with the numerical dissipation are significantly reduced in the high-resolution experiments. In more recent 3D simulations, \citet{Adrover:2020aap} and \citet{Fan:2020apj} pointed out that the reduction of the numerical dissipation is crucial to study the damping of LALOs.
\citet{Terradas:2013apj}, \citet{Luna:2016apj} and \citet{Zhang:2019apj} performed 2D numerical experiments of the prominence oscillation with spatial resolution up to $125$ and $156$ km, respectively. These numerical simulations showed significant damping which might be partly contributed by numerical diffusivity.
% of oscillations not associated with any physical mechanism. 
\citet{Liakh:2020aap} studied the convergence of a 2.5D experiment of LALOs excited in a magnetic flux rope. The authors performed an experiment with a spatial resolution of $60$ km and compared it to the one with a spatial resolution of $240$ km. They found that the periods of LALOs are consistent in the two experiments, although the damping time becomes longer in the higher resolution experiment. Thus, they concluded that in their experiments, the attenuation of LALOs is associated mainly with numerical dissipation.

In this work, we aim to understand the physical origin of the LALOs damping and the other processes usually hidden by the artificial dissipation in numerical experiments. To this end, we have performed 2D convergence experiments with progressively increasing spatial resolutions up to the highest value of 30 km.

This paper is organized as follows, in Sect. \ref{sec:numerical-model} the numerical model is described. In Sects. \ref{sec:results} and \ref{sec:study-damping-amplification} we explain the temporal evolution of the plasma under the disturbance directed along the field and compare the oscillatory parameters in the experiments with the different spatial resolution and the different prominence layers. In Sect. \ref{sec:discussion} we summarize the main results.

\section{Initial configuration}\label{sec:numerical-model}

\begin{figure*}[!ht]
	\centering
	\includegraphics[width=0.9\textwidth]{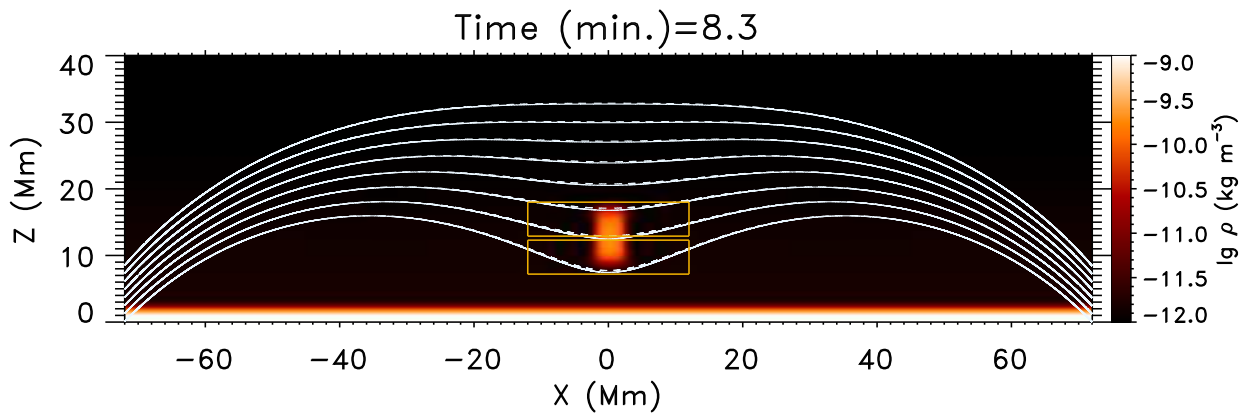}
	\caption{Density distribution and magnetic field lines at the central part of the computational domain after the mass loading and relaxation processes. The dashed lines denote the initial magnetic field prior to the mass loading. \label{fig:setup}}
\end{figure*}

We assume a 2D prominence model as shown in Fig. \ref{fig:setup}. The model is defined in the Cartesian coordinate system in the $xz$-plane, where $z$-axis corresponds to the vertical direction. The initial magnetic field is a potential configuration which contains a dipped part, suitable to support a prominence. It is composed of periodic arcades defined as:
\begin{eqnarray}\label{magnetic-field-component-x}
\frac{B_{x}}{B_{0}}&=&\cos(k_{1}(x-x_{0}))e^{-k_{1}(z-z_{0})}- \cos(k_{2}(x-x_{0}))e^{-k_{2}(z-z_{0})} \, , \\
\label{magnetic-field-component-z}
\frac{B_{z}}{B_{0}}&=&-\sin(k_{1}(x-x_{0}))e^{-k_{1}(z-z_{0})}+ \sin(k_{2}(x-x_{0}))e^{-k_{2}(z-z_{0})} \, ,
\end{eqnarray}
where $B_{0}=33$ G, $x_{0}=0\Mm$ and $z_{0}=-2\Mm$. We also set $k_{1}=\frac{\pi}{D}$ and $k_{2}=3k_{1}$ where $D=191.4\Mm$, that is the half-size of the numerical domain along the $x$-direction.
The superposition of the major and minor arcades form the magnetic field structure shown in Fig. \ref{fig:setup} with the dipped part centered at $x=0$ Mm and between $z=0$ and $25\Mm$. As shown in the figure, the curvature of the magnetic field lines decreases with height. For $z>27 \Mm$ the magnetic field lines change from concave-up to concave-down becoming ordinary loops. A detailed description of this magnetic configuration is given by \citet{Terradas:2013apj} and \citet{Luna:2016apj}. The magnetic field strength increases from $9.0$ to $12.5$ G for $z$ from $8$ to $25$ Mm. The magnetic field strength lies in the range of the values obtained from the magnetic field measurements in the solar prominences \citep{Leroy:1983solphys,Leroy:1984aa}.

We assume an initial atmosphere consisting of a stratified plasma in hydrostatic equilibrium, including the chromosphere, transition region (TR), and corona. The temperature profile is given by
\begin{equation}\label{eq:temperature-profile}
T(z)= T_{0} + \frac{1}{2} \left(T_\mathrm{c}-T_{0} \right) \left[1+ \tanh
\left( \frac{z-z_{c}}{W_z}\right) \right] \, . 
\end{equation}
We choose $T_\mathrm{c}=10^{6}$ K, $T_{0}=10^{4}$ K, $W_z=0.4\Mm$, and $z_{c}=3.6\Mm$. In this profile the temperature ranges from $T_ {ch}=10^{4}$ K at the base of the chromosphere to $T_{c}=10^{6}$ K in the corona. As the plasma is stratified in the vertical direction the density changes from $\rho=9\times 10^{-9}\ \mathrm{kg\ m^{-3} }$ in the chromosphere to $\rho=1.69\times 10^{-12}\ \mathrm{kg\ m^{-3} }$ at the base of the corona at the height $z_{c}=3.6$ Mm. 

We  numerically solve ideal magnetohydrodynamic (MHD) equations using the MANCHA3D code \citep{Khomenko:2008solphys, Felipe:2010apj, Khomenko:2012apj}. The governing equations of mass, momentum, internal energy, induction equation, and the corresponding source terms are described in \cite{Felipe:2010apj}. The computation domain consists of a box of $384 \times 108\Mm$ size. In order to study the influence of the numerical diffusivity on the prominence oscillations, we use four spatial resolutions with $\Delta =240,\ 120,\ 60\, $ and $30$ km from the coarse to the fine grids. We assume a periodic condition at the left and right boundaries. At the bottom boundary, we apply the current-free condition for the magnetic field \citep[see, e.g.,][]{Luna:2016apj} and symmetric condition for the temperature and pressure. At the bottom boundary, the density is fixed. At the top boundary, the zero-gradient condition is applied to all the variables except for $B_{x}$. We impose that $B_{x}$ is antisymmetric. 
\begin{figure*}[!h]
	\centering
	\includegraphics[width=0.95\textwidth]{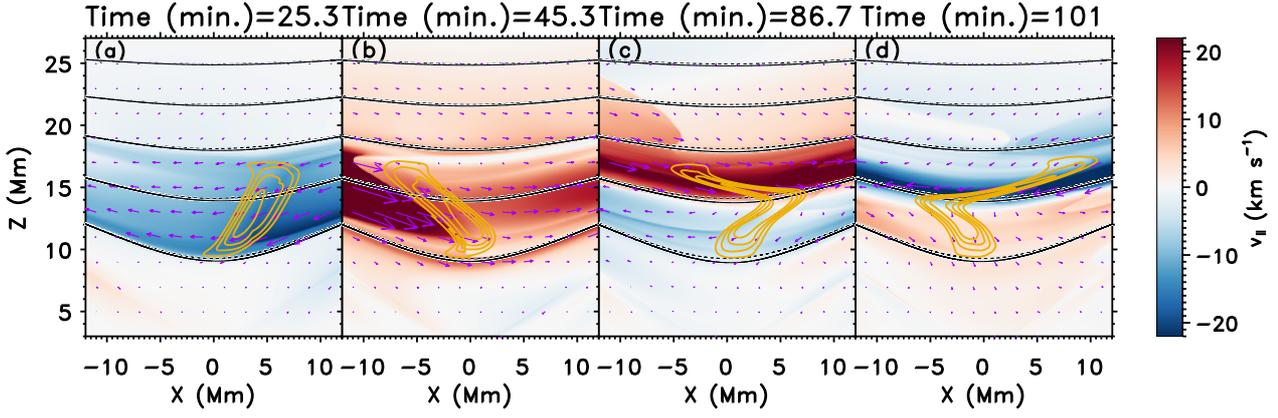}
	\caption{Temporal evolution of $v_{\parallel}$, magnetic field, and the prominence mass after the perturbation. Dashed black lines denote unperturbed magnetic field lines; solid black lines are the actual magnetic field lines at a given moment; orange lines are density isocontours, and purple arrows are the velocity field. Spatial resolution: $\Delta =30$ km. \label{fig:evolution}}
\end{figure*}

In order to add the prominence mass to the dipped region of the magnetic field, we use an approach of the artificial mass loading described by \citet{Liakh:2020aap}. We use the source term in the continuity equation to increase the density in the dipped region of the magnetic field. The density source term distribution is a Gaussian function (see Eq. (9) from \citet{Liakh:2020aap} work) centered at $(x,z)=(0,11.8)\Mm$. The mass loading starts at $t=0\secs$ and ends at $t=100 \secs$. The resulting prominence has a density of $110$ times the density of the initial corona with the dimensions of $5$ and $9$ Mm in the horizontal and vertical directions, respectively.

During the first $8.3\mins$, we use intensive artificial damping in order to minimize the undesirable motions caused by the response of the magnetic field to the mass loading process. Figure \ref{fig:setup} shows the prominence after this relaxation phase. The dashed lines denote the initial magnetic field. As we can see, the magnetic field lines are slightly elongated downwards due to the heavy prominence mass. The prominence itself also has some deformation, becoming more compressed toward the center of the dip, and it slightly drops down with respect to the initial height. At the end of the relaxation process, the system is close to static equilibrium.

Similarly to our previous work \citep{Liakh:2020aap}, we trigger oscillation after the relaxation phase by applying an external force over the prominence. This force is incorporated as a source term in the momentum equations as
\begin{eqnarray}\label{eq:momentum-sourcex}
S_{mx}=\frac{\rho v_{pert}B_{x}}{t_{pert}B}\exp\left({-\frac{(x-x_{pert})^4}{\sigma_x^4}-\frac{(z-z_{pert})^4}{\sigma_z^4}}\right) \, , \\ \label{eq:momentum-sourcez}
S_{mz}=\frac{\rho v_{pert}B_{z}}{t_{pert}B}\exp\left({-\frac{(x-x_{pert})^4}{\sigma_x^4}-\frac{(z-z_{pert})^4}{\sigma_z^4}}\right) \, ,
\end{eqnarray}
where $t_{pert}=10\secs$ is the duration of the disturbance that starts at $t=8.3 \mins$. The parameters $\sigma_{x}=\sigma_{z}=12\Mm$ are the half-sizes of the perturbed region in the horizontal and vertical directions centered at $(x_{pert},z_{pert})=(0,12) \Mm$. The force given by Eqs (\ref{eq:momentum-sourcex}) and (\ref{eq:momentum-sourcez}) is directed along the local magnetic field in contrast to our previous work \citep{Liakh:2020aap} where the perturbation is purely horizontal or vertical. 
The maximum velocity of the perturbation is $v_{pert}=22\kms$, which is in the range of the observed amplitudes of LAOs \citep[see review by][]{Arregui:2018spr}. In order to study the long-term evolution of prominence motions, we run the experiments during $275 \mins$ of physical time.

\section{Influence of spatial resolution on prominence dynamics}\label{sec:results}

\begin{figure*}[!h]
	\centering
	\includegraphics[width=0.9\textwidth]{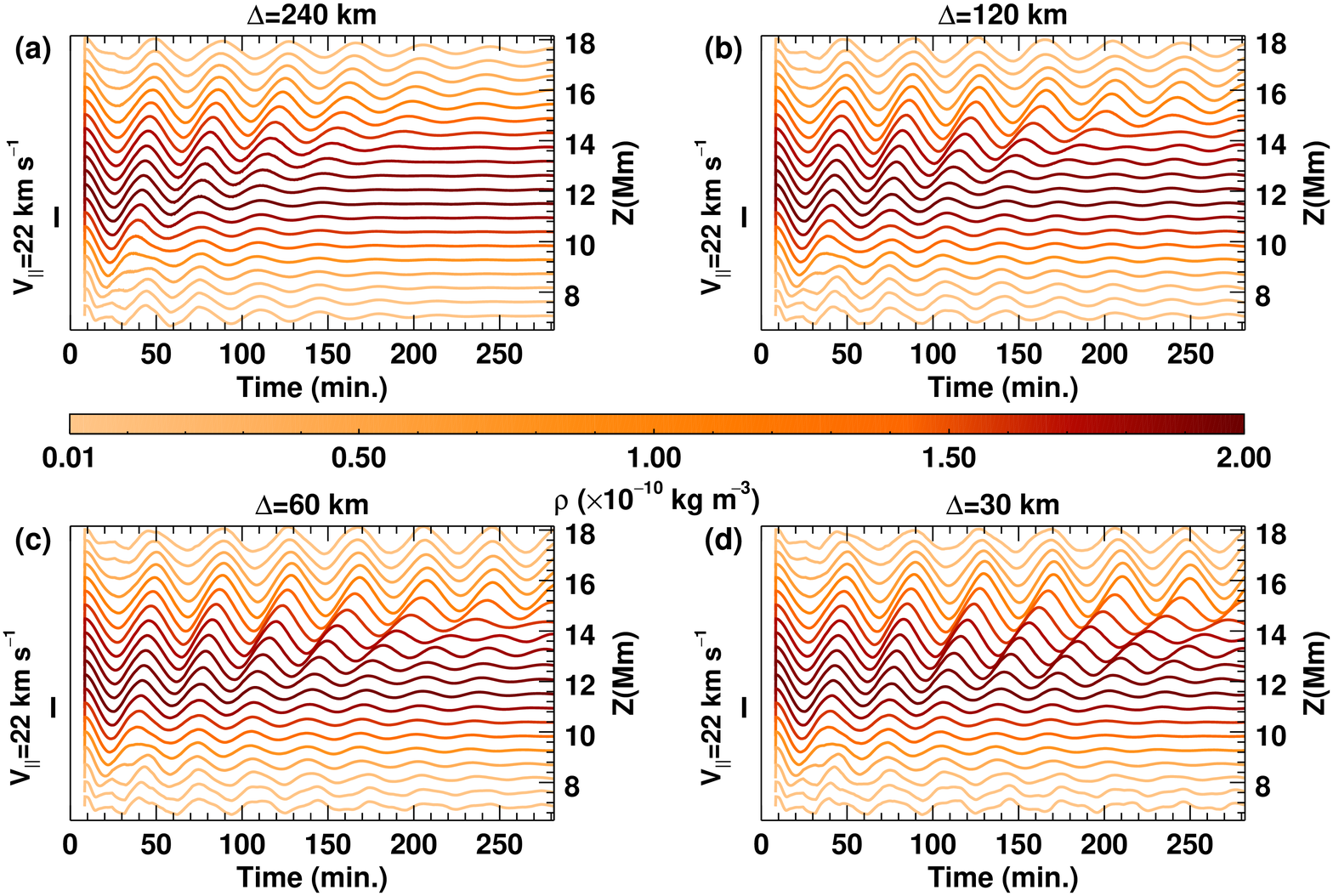}
	\caption{Temporal evolution of $v_{\parallel}$ at the center of mass at the selected field lines. The color bar denotes the maximum initial density at each field line. The left vertical axis indicates the velocity amplitude scale. The right vertical axis denotes the vertical location of the dips of selected magnetic field lines. \label{fig:longitudinal-velocity}}
\end{figure*}
In this work, we study the influence of the numerical diffusivity in the LAOs by progressively increasing the spatial resolution as described in Sect. \ref{sec:introduction}. Figure \ref{fig:evolution} shows an example of the temporal evolution after the perturbation for $\Delta=30$ km. This kind of evolution is representative of all our numerical experiments. After the initial disturbance at $t=8.3 \mins$, the prominence moves to the right along the magnetic field and reaches its maximum initial displacement at $t=17 \mins$, approximately (Fig. \ref{fig:evolution}a). 
We see that the maximum horizontal displacement is large at the upper prominence part, whereas the bottom part is only slightly displaced from the equilibrium position.
The reason is that, as the initial perturbation launches the prominence mass along the field lines with similar velocities, the plasma reaches different horizontal locations since the field lines have different curvatures.
Figure \ref{fig:evolution}b shows the prominence at $t=45.3 \mins$. At this time, the prominence has reached its maximum displacement on the left side and begins to move again toward the right. The top part of the prominence slightly delays from the rest of the prominence body, suggesting a longer oscillation period. 
Figures \ref{fig:evolution}c and \ref{fig:evolution}d show the prominence evolution after several cycles of oscillations. From the velocity field displayed by the purple arrows, we can see that the prominence layers oscillate out of phase producing counterstreaming flows. The counterstreaming flows are commonly observed in the filaments \citep{Schmieder:1991aap, Zirker:1998nat, Lin:2005solphys}. \citet{chen_imaging_2014} claimed that these counterstreaming motions are associated with oscillations.
The oscillation period depends on the physical conditions of the magnetic field line where the prominence plasma resides. This means that the period can vary slightly for the neighboring field lines. This explains a phase difference between the motions of the prominence layers, shown in Fig. \ref{fig:evolution}c. Besides, these alternate motions can also be associated with processes of evaporation and condensation in prominences \citep{Zhou:2020nat} or shock downstreams produced by jets \citep{luna_large-amplitude_2021}.
 
In order to study the prominence plasma oscillations, we analyze the bulk motions of the plasma in the two possible polarizations: transverse and longitudinal to the magnetic field. Following our previous works \citep{Luna:2016apj, Liakh:2020aap} we compute the velocities of the center of mass of independent field lines using Eqs. (5) and (6) from \citet{Luna:2016apj}. The parameters $v_{\parallel}$ and $v_{\perp}$ are the longitudinal and transverse velocities of the center of mass of each field line. First, we select 20 equally spaced field lines that permeate the prominence body at the height from $7.2$ Mm up to $17.6$ Mm. Then, we start integrating the field lines in the chromosphere, where the magnetic field remains unchanged according to the line-tying condition. As our fluid is perfectly conducting, the frozen-in condition is fulfilled. This allows us to advect the motions in the same field line at each time moment. We selected an identical set of the field lines for each numerical experiment in order to compare the resulting velocities.

Figure \ref{fig:longitudinal-velocity} shows the temporal evolution of $v_\parallel$ for the four numerical experiments with different spatial resolutions. The global behavior of the prominence is relatively similar in each of the experiments. After the initial perturbation at $t=8.3 \mins$, the longitudinal velocity increases in all the lines. The longitudinal velocity reaches the highest value, $v_{\parallel}=22 \kms$, in the densest prominence part at height $z=11.8$ Mm. At $t=15-35 \mins$, we see a certain shift in the signal. This phase shift is related to the different periods associated with the different field lines. On the one hand, we see that the oscillations are less attenuated with increasing the spatial resolution, although the difference between the cases $\Delta=60$ and $30$ km is less pronounced. On the other hand, the behavior of the damping changed with height. This is in agreement with the visual impression from Fig. \ref{fig:evolution}, discussed above. Oscillations at the bottom and central part are strongly damped in all the experiments. At the height of $13-17$ Mm, the oscillations with weaker attenuation last for a longer time. In the highest-resolution simulation, the upper prominence part keeps oscillating with a significant amplitude even at the final stage of the numerical experiment.

%%%%%%%%%%%%%%%%%%%%%%%%
\subsection{Influence of spatial resolution on the period}\label{subsec:influence-on-periods}

\begin{figure}[!t]
	\centering\includegraphics[width=0.45\textwidth]{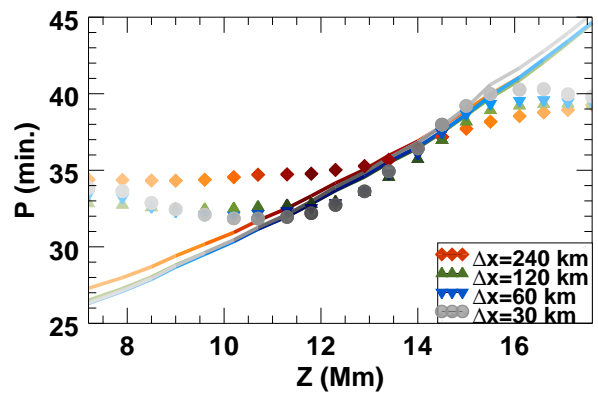}
	\caption{Periods as a function of height of the magnetic dips. The symbols denote the longitudinal period obtained from the numerical experiments, and the solid lines are the periods predicted by the pendulum model. Different colors and symbols correspond to the experiments with different spatial resolutions. The color gradation from dark to light corresponds to decrease of the density contrast. \label{fig:periods}}
\end{figure}
Figure \ref{fig:periods} shows the periods of the longitudinal oscillations for each of the 20 selected field lines and each numerical experiment with different spatial resolutions. The period is more or less uniform under $10.2$ Mm and above $16.1$ Mm. In contrast, for heights $z=10.2-16.1$ Mm, we see an increase of the period with the height of the magnetic dip. These heights correspond to the prominence region with the highest density contrast. One can observe in this figure that the slope of the curves depends on the spatial resolution. The steepest slope corresponds to the experiment with the highest spatial resolution ($\Delta=30$ km). In contrast, for $\Delta=240$ km, the period is more uniform with a smoother variation with height.

Under prominence conditions, the main restoring force of these oscillations is the solar gravity projected along the magnetic field \citep[see e.g.,][]{Luna:2012apjl, Luna:2012apj, Luna:2016apj, Zhou:2018apj, Liakh:2020aap, Fan:2020apj}. \citet{Luna:2012apjl} suggested that the period of the longitudinal oscillations depends only on the radius of curvature of the magnetic dip, $R_{c}$, as
\begin{equation}\label{pendulum-period}
P=2\pi\sqrt{\frac{R_{c}}{g}} \, ,
\end{equation}
where $g=274 \, \mathrm{m\, s^{-2}}$ is the solar gravitational acceleration. 
In our magnetic configuration, the curvature of the magnetic field lines in the prominence area decreases with height. This implies that the period of the different field lines increases with height in agreement with Fig. \ref{fig:longitudinal-velocity}. 
The radius of curvature of the dips of the field lines changes with time in response to plasma motion. In order to compare the results of the simulations with the pendulum model, we compute the time-averaged radius of curvature. The values obtained are $R_{c}=17.3-51.1$ Mm for the selected field lines from the bottom to the top of the prominence. With Eq. \eqref{pendulum-period} we computed the theoretical period shown in Fig. \ref{fig:periods} as a solid line. From the figure, we see that for the experiments with $\Delta=30,60$ and $120$ km, we have a good agreement between the numerical results and the pendulum model at heights $z=10.2-16.1$ Mm, where the body of the prominence is located. We also observe that even though all three resolutions have a good agreement, the agreement is slightly better for the finest resolution. In contrast, in the coarse case with $\Delta=240$ km, there is only agreement at the central part of the prominence. This indicates a strong influence of the numerical diffusivity on the period of oscillations and that it is necessary to reach a certain spatial resolution to have an agreement with the pendulum model. Using 3D numerical simulation, \citet{Zhou:2018apj} and \citet{Fan:2020apj} concluded that in their experiments a disagreement with pendulum model is about of $10$ \%. We suggest that high-resolution experiments can reduce this discrepancy.

Figure \ref{fig:periods} shows a clear deviation from the pendulum model below the prominence, $z<10.2$ Mm and above it, $z>16.1$ Mm, even for high spatial resolution. This behavior suggests an existence of a physical reason for the deviation. Previous works \citep[such as ][]{Zhou:2018apj, Zhang:2019apj, Liakh:2020aap, Fan:2020apj} have also found discrepancies with the pendulum model. 
\citet{Zhou:2018apj} and \citet{Zhang:2019apj} found that if the gravity to Lorentz force ratio is close to unity, the heavy prominence plasma can significantly deform the magnetic field lines.
They defined a dimensionless parameter as the ratio between the weight of the thread to the magnetic pressure as 
\begin{equation}\label{delta-zhou}
 \delta =\frac{2 \rho g l }{B^2/2\, \mu_0} \, ,
\end{equation}
where $l$ is the half-length of the prominence thread. The authors argued that if the parameter $\delta$ is close to unity, the weight of the prominence changes the field geometry dynamically.
In this situation, the actual trajectory of the plasma does not coincide with the field lines, and the trajectory of the plasma has a larger radius of curvature than the one corresponding to an unperturbed magnetic field line \citep[see Fig. 9 from][]{Zhang:2019apj}. 
Thus, the resulting period of LALOs can be longer than predicted by the pendulum model. In our model, the magnetic field at the dipped part increases with $z$, from $9$ G at the base of the prominence up to $12.5$ G at its top. Using the parameters of our model, we obtain that the maximum value $\delta=0.7$ reached at around $z=11.8\Mm$, that is around the densest region of the prominence. In that region, we have obtained a good agreement with the pendulum model. For the regions below and above the prominence, $\delta$ is much smaller. We conclude that the weakness of the magnetic field cannot explain the discrepancies between the pendulum model and our simulations.
However, we have found that the interaction between the longitudinal motion of the prominence and the magnetic structure is not negligible. When the prominence moves, the magnetic field structure changes considerably. These perturbations of the magnetic field could be transmitted to the rest of the field structure. This contributes to the damping of the LALOs due to wave leakage, as we discuss below in Sect. \ref{sec:study-damping-amplification}. 

\citet{Luna:2012apj} showed that the restoring force of LALOs is a combination of the gravity projected along the magnetic field and gas pressure gradient. The relative importance of both restoring forces depends on the radius of curvature of the field lines with respect to a characteristic radius $R_{lim}$. In a situation where $R_{c} \ll R_{lim}$, the gravity dominates, and the pendulum model is valid. In contrast, for $R_{c} \gtrsim R_{lim}$ the gas pressure term dominates. The radius $R_{lim}$ depends on several parameters of the structure
\begin{equation}\label{radii-ratio}
R_{lim}=\frac{l(L-l)\chi g}{c_{sc}^{2}} \, ,
\end{equation}
where $c_{sc}$ is coronal sound speed, $L$ is half-length of the field line, $\chi$ is the density contrast between the prominence and the ambient corona. In our situation, below the prominence ($z<10.2$ Mm), the ratio $R_{c}/R_{lim}>0.2$, and it increases for smaller values of $z$. The main reason is that the density contrast $\chi$ is close to $1$ in that region. This indicates that the contribution of the gas pressure to the restoring force is important, and the pendulum model is no longer valid.
Similarly, for $z\ge16.1$ Mm, both the curvature of the field lines and the density contrast decrease with height. Thus $R_{c}/R_{lim}$ increases in these field lines from $0.4$ up to $5.9$ for higher values of $z$. This explains why the period of the longitudinal oscillations deviates from the pendulum period for locations below and above the prominence. In turn, in the central region, $z=10.2-16.1$ Mm, the ratio is around of $0.1$.

\begin{figure}[!ht]
	\centering\includegraphics[width=0.45\textwidth]{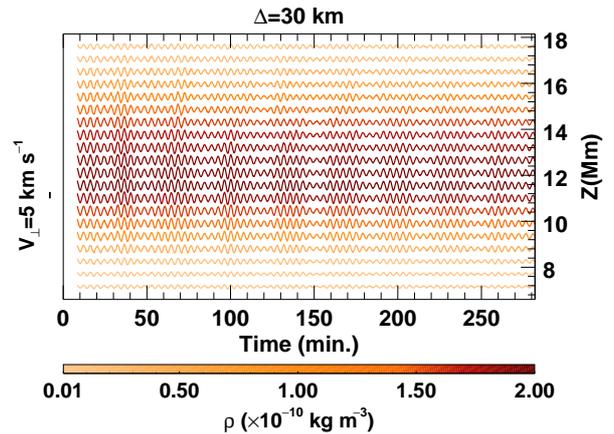}
	\caption{Temporal evolution of $v_{\perp}$ at the center of mass of the selected field lines. The color bar denotes the maximum initial density at each field line. The left vertical axis indicates the velocity amplitude scale. The right axis denotes the height of the dips of the field lines. \label{fig:transverse-velocity}}
\end{figure}
The prominence is also subject to transverse oscillations. Similarly to the longitudinal velocity, we obtained temporal evolution of $v_{\perp}$ using Eq. (6) from \citet{Luna:2016apj} for each selected field line. Figure \ref{fig:transverse-velocity} shows the temporal evolution for the $\Delta = 30$ km case. We only show the case with the finest spatial resolution because all experiments have a very similar temporal evolution of the transverse velocity. 
From the figure, we see that the transverse motions are synchronized in all the field lines with the same oscillatory period, $P=4\mins$. This uniformity of the oscillations shows that the motion is related to a global normal mode of the field structure.
The oscillation in $v_{\perp}$ is not harmonic, and the amplitude is modulated every $30\mins$ approximately. This modulation is also synchronized in all the field lines shown in the figure. 
The amplitude of the transverse oscillations is equal to $5\kms$ that is much smaller than for the longitudinal motions.
This modulation of the amplitude is probably related to the motion of the prominence mass. The global motion of the plasma changes the characteristics of magnetic structure periodically, resulting in a modulation of the oscillation. The signal modulation of the transverse velocity can also indicate the longitudinal to transverse mode conversion. It can be a subject for future work.
During the simulation time, the transverse oscillations show nearly no significant damping in the four numerical experiments with different $\Delta$, indicating that numerical dissipation does not affect this kind of motion. 
%In addition, Figure \ref{fig:transverse-velocity} shows that there is no evidence of the longitudinal to transverse mode conversion that could explain the strong damping of the longitudinal oscillations in the central prominence part. We do not see an increase of the amplitude in the signal of $v_{\perp}$ in the field lines with $z=7.6-12.3$ Mm associated with strong attenuation of $v_{\parallel}$.
%\textbf{As we discuss in Sec XXX the damping is associated mainly to the emission of fast MHD waves. These waves perturbs the transverse component of the velocity. From Figure \ref{fig:transverse-velocity} we do not see a clear signature of transfer of energy from longitudinal modes to transverse waves. However, it is necessary and in-depth study that is out of the scope of the current research.}

\subsection{Influence of spatial resolution on the damping}\label{subsec:influence-resolution-damping}

\begin{figure}[!t]
	\centering
	\includegraphics[width=0.45\textwidth]{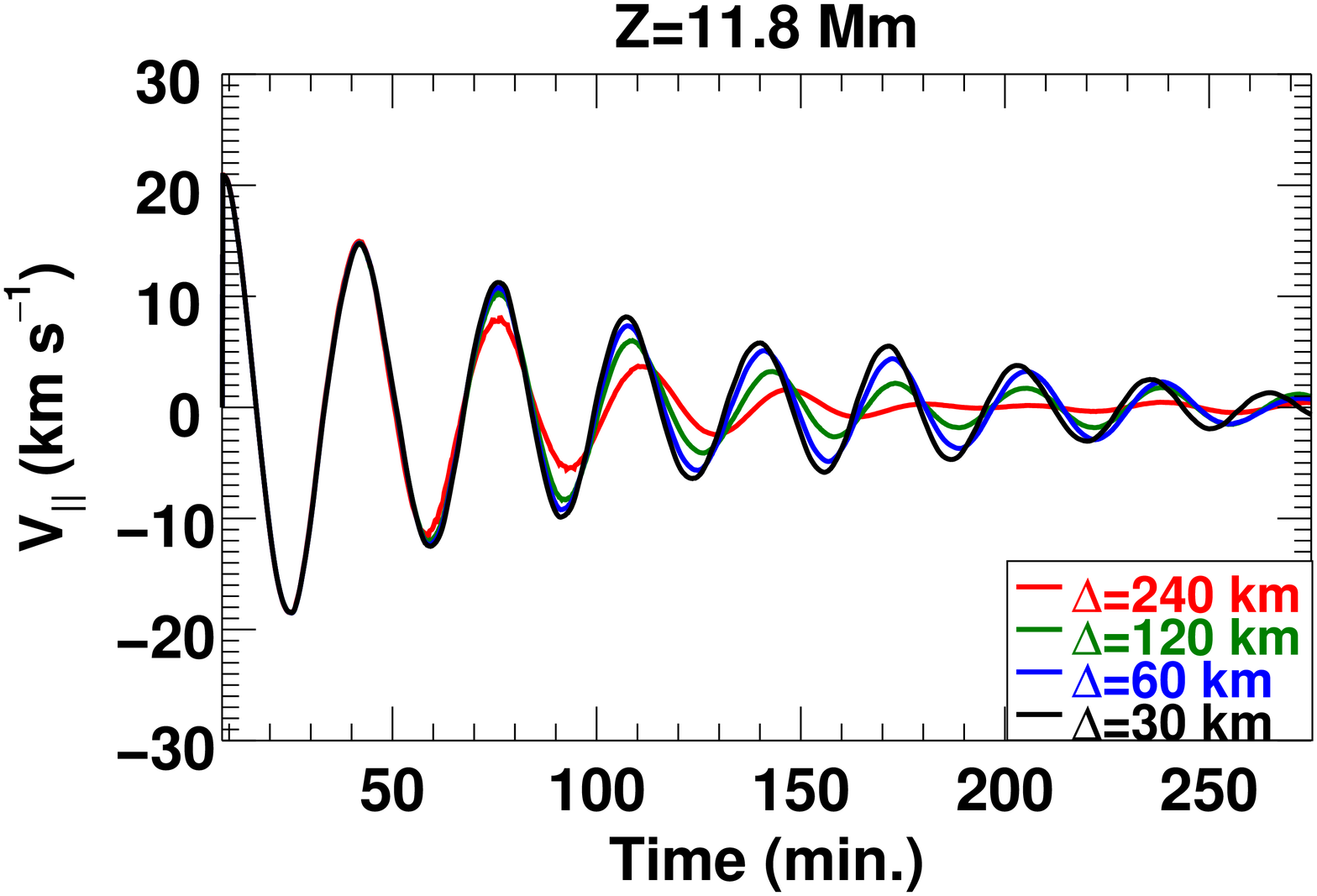}
	\includegraphics[width=0.45\textwidth]{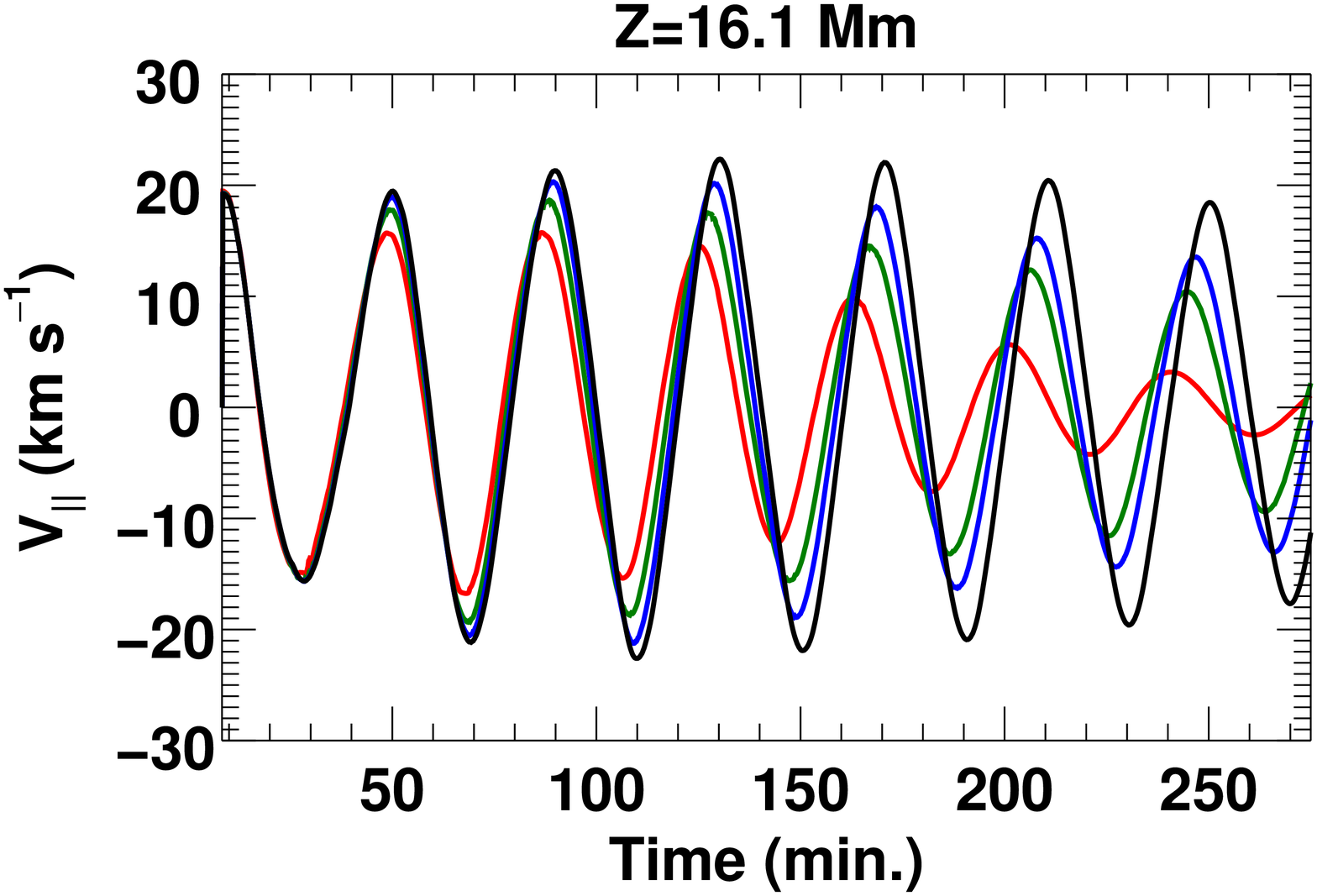}
	\caption{Temporal evolution of $v_{\parallel}$ at the center of mass of the selected field lines. Top: field line close to the prominence center; bottom: field line at the top part of the prominence. The colors stand for the experiments with different spatial resolutions. \label{fig:comparison-center-top}}
\end{figure}

\begin{figure}[!t]
	\centering
	\includegraphics[width=0.45\textwidth]{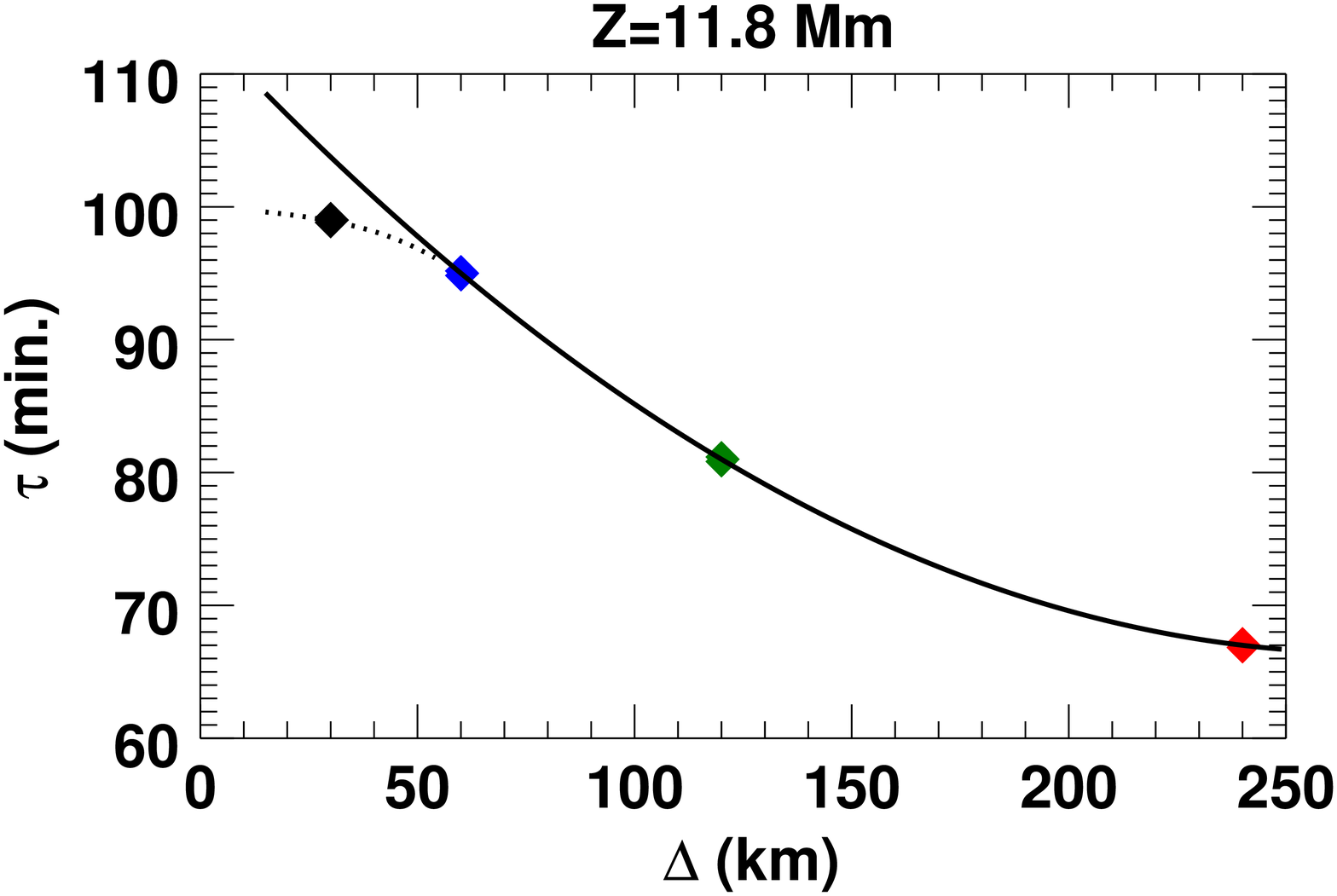}
	\includegraphics[width=0.45\textwidth]{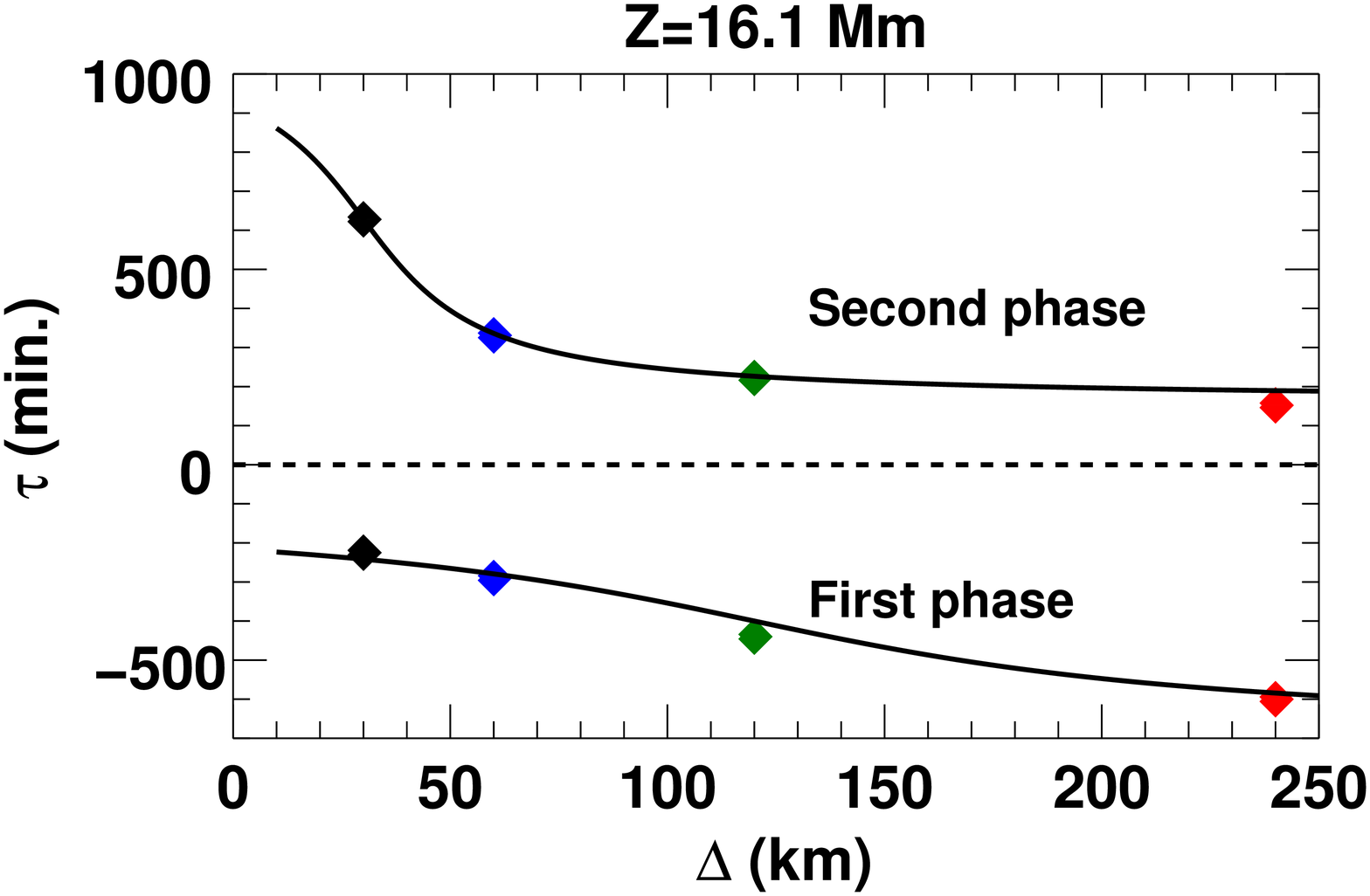}	\caption{Top: damping time of oscillations at height $11.8$ Mm as a function of the spatial resolution $\Delta$. The solid line denotes the quadratic dependence, and the dashed line shows fitting with an arctangent function. Bottom: the same for the height $16.1$ Mm. The solid lines denote fitting with an arctangent function. \label{fig:damping-center-top}}
\end{figure}

In order to study in detail the damping in the experiments with different $\Delta$, we consider longitudinal velocity in two field lines with the dips at $z=11.8$ Mm (central part of the prominence) and $z=16.1$ Mm (top part). The $v_\parallel$ in the central and top field lines are shown in Fig. \ref{fig:comparison-center-top}. The upper panel of Fig. \ref{fig:comparison-center-top} shows that the attenuation is strongest in the experiment with the coarsest resolution ($\Delta=240$ km). The oscillation signal is almost completely damped after $150 \mins$. In contrast, the experiment with $\Delta=30$ km is where observed the weakest attenuation. Comparing all the cases, we see that the damping time increases for increasing spatial resolution (i.e., decreasing $\Delta$). The curves for $\Delta=60$ and $30$ km are relatively close to each other. This indicates that the damping time is close to the saturation and that the high-resolution experiment shows indications for the physical damping not associated with numerical diffusivity. In order to quantify the damping time we fit the signal of $v_{\parallel}$ by decaying sinusoidal function such as $v_{\parallel}=V_{0}e^{-t/\tau_{d}}\sin(2\pi t/P + \phi)$. The values of $\tau_{d}$ obtained from the best fit to $v_{\parallel}$ are shown in Fig. \ref{fig:damping-center-top} (top panel) as a function of spatial resolution, $\Delta$. 
We observe that the damping time increases when $\Delta$ decreases for experiments with $\Delta =240, 120,$ and $60$ km. In these three experiments, the damping time shows a quadratic dependence on $\Delta$ denoted by the solid line in the top panel of Fig. \ref{fig:damping-center-top}.
The trend marked by the solid line seems to indicate the damping time of about $\tau_{d} =110\mins$ as $\Delta$ approaches zero. However, the damping time in the experiment with $\Delta =30$ km deviates from this quadratic trend and has a similar value to the one in the $\Delta = 60$ km experiment. We fitted an arctangent function to all four values of $\tau_{d}$, including the one in $\Delta =30$ km experiment. This new fit is shown as a dashed line at the top panel of Fig. \ref{fig:damping-center-top}. The trend, including the highest resolution case, deviates from the quadratic trend as the curve becomes flatter. This suggests that $\tau_{d}$ may saturate for decreasing values of $\Delta$ showing the non-numerical (i.e., physical) origin of this damping. According to the dashed line, $\tau_{d} \rightarrow 100\mins$ as $\Delta$ approaches zero.

We repeated the same analysis for the magnetic field line located at $z=16.1$ Mm, which is at the top of the prominence. The results are shown at the bottom panel of Fig. \ref{fig:comparison-center-top}. The temporal evolution of the longitudinal velocity differs much from the one at $z=11.8\Mm$. We can clearly distinguish two stages in temporal evolution. In the first stage, the velocity is amplified after the initial perturbation. The amplification is more significant and extended in time in the simulations with higher spatial resolution. For the case of $\Delta=30$ km, the amplitude increases from $v_{\parallel}=19.5\kms$ up to $23 \kms$ during $130 \mins$. This amplification can be well traced in the snapshots of temporal evolution shown in Fig. \ref{fig:evolution}d above. The oscillations are attenuated in the bottom and central parts of the prominence. However, at heights $z>15$ Mm, the displacement is not only comparable but even larger than the initial displacement shown in Fig. \ref{fig:evolution}a.
During the second stage, the oscillations begin to decay slowly. At the end of simulated time ($t=275 \mins$), the motions are almost completely damped for the experiments with $\Delta=240, 120$ km. In the higher-resolution simulations, $\Delta=30$ and $60$ km, the upper prominence part oscillates undamped for a longer time. In order to quantify the damping time for these complex oscillations, we split the velocity signal into two parts associated with the amplification and damping. We then fit $v_{\parallel}$ with the decaying and amplifying sinusoidal functions. For the amplification stage, we assume the characteristic time, $\tau_{A}$, to be negative, and for the decaying stage, the damping time, $\tau_{D}$, positive. From the best fit to the longitudinal velocity we obtained the values of $\tau_{A}$ in the range $\left[ -600, -225\right] \mins$, and those of $\tau_{D}$ in the range of $\left[152, 628\right] \mins$. The higher values of both parameters correspond to better resolutions. The results of this analysis are shown at the bottom panel of Fig. $\ref{fig:damping-center-top}$. The characteristic time of the amplification, $\tau_{A}$, differs significantly in the high- and low-resolution experiments. We can observe that, in the experiment with $\Delta=30$ km, $\tau_{A}=-225 \mins$, that is the most significant amplification. In addition, the bottom panel of Fig. \ref{fig:comparison-center-top} also shows that this amplification stage is more extended in time for the finer spatial resolutions. The damping time in the second phase reaches the value of $\tau_{D}=628\mins$ for the simulation with $\Delta=30$ km, indicating the weakest damping. The amplification stage is not so relevant for the experiment with $\Delta=240$ km. For this experiment, the amplification time $\tau_{A}=-600 \mins$ is the longest, and the oscillations have almost constant amplitudes during the first $67.8\mins$. We performed a fit to both $\tau_{A}$ and $\tau_{D}$ using an arctangent function, allowing us to find the trend for both parameters (Fig. \ref{fig:damping-center-top}, solid lines at the bottom panel). With these trends, we can roughly estimate that the real physical amplification for our system would have a characteristic time of $-200$ minutes, whereas the damping time would be longer than $900$ minutes. These results may indicate that our highest-resolution experiment is about to resolve a physical mechanism for these effects.

\section{Physical reasons for the amplification and damping of oscillations}\label{sec:study-damping-amplification}

As we have shown in Sect. \ref{subsec:influence-resolution-damping} above, oscillations at the bottom part of the prominence damp quickly, and those at the top part are initially amplified and damped later. We have also seen that these phenomena may have a physical origin and are not related to numerical dissipation. In this section, we perform a detailed analysis to shed light on the possible physical mechanisms that produce these effects.

%%%%%%%%%%%%%%%%%% BEGINNING OF THE ANALYSIS OF THE ENERGY IN A FINITE VOLUME %%%%%%%%%%%%%%%%%%%%%
First, we study the contributions of the different forces to the energy balance in two regions of the prominence defined by the orange rectangles in Fig. \ref{fig:setup}.
\begin{figure*}[!ht]
	\centering
	\includegraphics[width=0.95\textwidth]{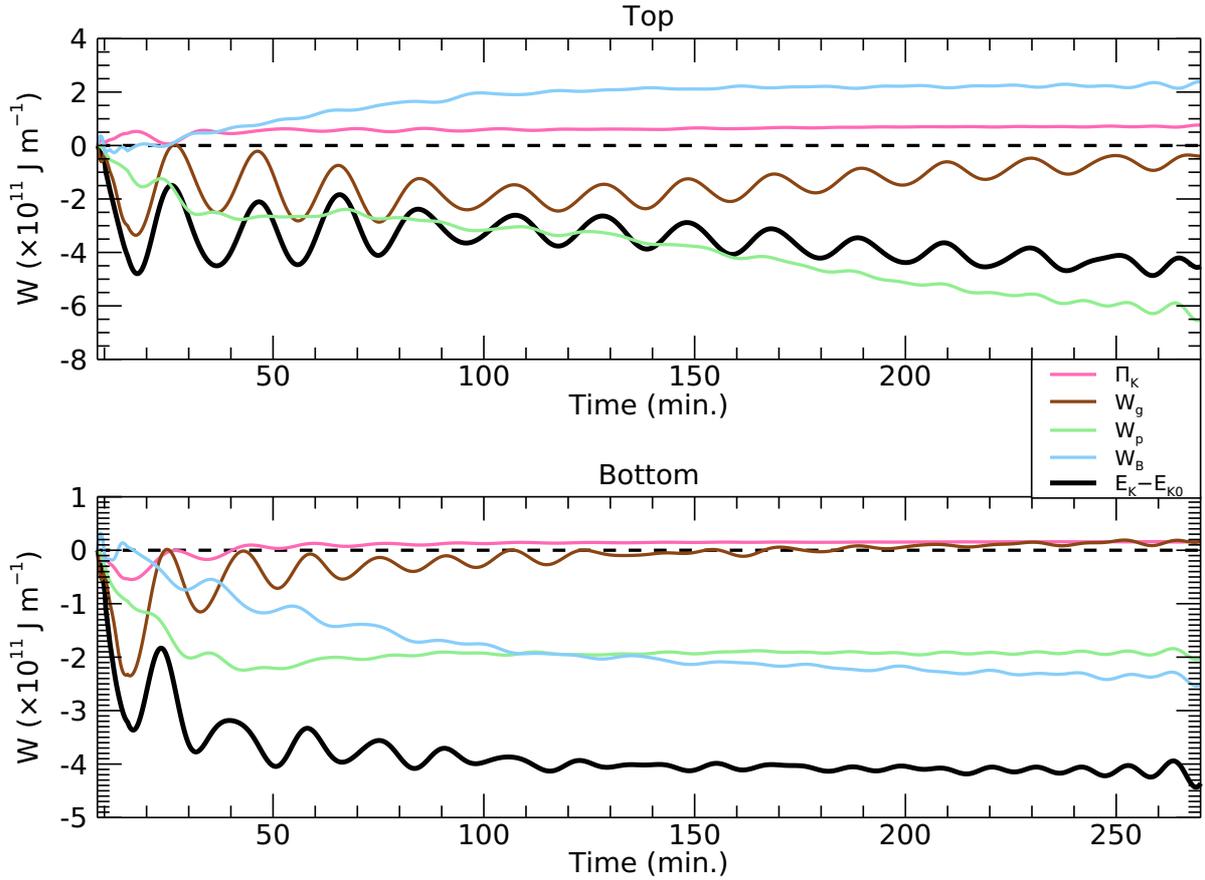}
	\caption{Work done by the gravity force (brown line), by the gas pressure force (light green line), by the Lorentz force (light blue line), and the kinetic energy flow (pink line) integrated in two rectangles shown by the orange lines in Fig. \ref{fig:setup}. The black lines denotes a total kinetic energy variation with respect to its value at $t=8.5\mins$.\label{fig:time-integral}}
\end{figure*}
Equations (\ref{gravity-force-work-tot}--\ref{lorentz-force-work-tot}) show the temporal derivatives of the works done by the different forces in the region, which we denote as $\Sigma$. In addition, Eq. (\ref{kinetic-energy-flow}) gives the kinetic energy flow along the boundaries of the region, $\partial \Sigma$. These equations are:
\begin{eqnarray}\label{gravity-force-work-tot}
	\overbigdot{W}_{g}&=&\iint_{\Sigma} \rho \, \mathbf{g}\mathbf{\cdot v} \, d\sigma\, , \\
	\label{gas-pressure-force-work-tot}
	\overbigdot{W}_{p}&=&-\iint_{\Sigma} \mathbf{v \cdot} \nabla p  \, d\sigma\, , \\
	\label{lorentz-force-work-tot}
	\overbigdot{W}_{B}&=&\iint_{\Sigma} \frac{1}{\mu_0}\mathbf{\left(\nabla \times B \right)\times \mathbf{B}} d\sigma\, , \\
	\label{kinetic-energy-flow}
	\overbigdot{\Pi}_{K}&=&-\oint_{\partial \Sigma} \frac{1}{2}\rho v^2 \mathbf{v\cdot n}\ dl\ \, ,
\end{eqnarray}
where $\mathbf{n}$ is the vector normal to the region boundaries. The combination of the four terms is equal to the temporal derivative of the kinetic energy integrated in the region $\Sigma$, $\overbigdot{E}_{K}$. Figure \ref{fig:time-integral} shows the temporal evolution of the terms given by Eqs. (\ref{gravity-force-work-tot}--\ref{kinetic-energy-flow}). All these magnitudes are shown with respect to their values at $t=8.5\mins$. The top and bottom panels of the figure show the quantities computed in the top and bottom region shown in Fig. \ref{fig:setup}, respectively.

Figure \ref{fig:time-integral} shows a clear difference in the evolution of the plasma in both regions of the prominence. At the top of the structure, the kinetic energy decreases rapidly right after the initial perturbation. The curve oscillates around an average value that seems to increase slightly from $30$ to $100\mins$, approximately. The amplification of the oscillations at the upper part of the prominence does not result in a significant increase of the kinetic energy because the region of integration is larger than the prominence size and includes the nonamplified regions. After $t=100\mins$ the kinetic energy decreases. Figure \ref{fig:time-integral} (top panel) shows that the gas pressure work is negative and that it makes the largest contribution to the kinetic energy losses after $t=100\mins$. In contrast, the work done by the Lorentz force in the first $100\mins$ is positive, indicating that it accelerates the prominence plasma. In turn, the kinetic energy inflow through the boundaries is relatively small compared to the other terms.

\begin{figure}[!ht]
	\centering
	\includegraphics[width=0.5\textwidth]{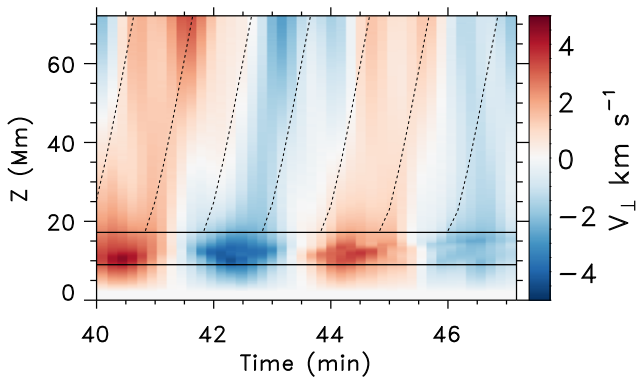}
	\caption{Time-distance diagram of the transverse velocity, $v_{\perp}$, along the vertical direction at horizontal position $x=0$ Mm. Black horizontal lines denote the position of the prominence. Dashed lines denote the position of a hypothetical wave front propagating at a local Alfv\'en speed. \label{fig:time-distance}}
\end{figure}

At the bottom part of the prominence (Fig. \ref{fig:time-integral}, bottom panel), the black line shows a substantial decrease in the first $100\mins$ coinciding with the significant attenuation of the oscillations in this region. The bottom panel shows that the main contribution to the kinetic energy losses is associated with the work of the gas pressure and the Lorentz force. This can indicate the generation of fast MHD waves, which are lately emitted, contributing to the damping of the oscillations, as already suggested by \citet{Zhang:2019apj}. These fast waves are associated with the gas pressure and the Lorentz forces. Additional evidence for the wave leakage is shown in Fig. \ref{fig:time-distance}. This figure shows the time-distance diagram of the transverse velocity, $v_{\perp}$, along the axis, $x=0$ Mm, during $10\mins$. The fast MHD waves produce perturbations of the $v_{\perp}$ field. The figure shows the waves that travel upwards with a pattern of alternating positive and negative values of $v_{\perp}$. The inclination of the ridges is indicative of the speed of waves.
Since we performed the experiments in the low-$\beta$ regime, the speed of the fast waves is approximately the Alfv\'en speed, $v_{A}$. The dashed lines represent locations of a wave front that would propagate at a local Alfv\'en speed. We can see that the waves emitted from the prominence propagate with speed similar to the Alfv\'en speed. Figures \ref{fig:time-integral} and \ref{fig:time-distance} indicate that an important portion of the energy of the prominence oscillation is emitted in the form of fast MHD waves.

\begin{figure}[!ht]
	\centering
	\includegraphics[width=0.45\textwidth]{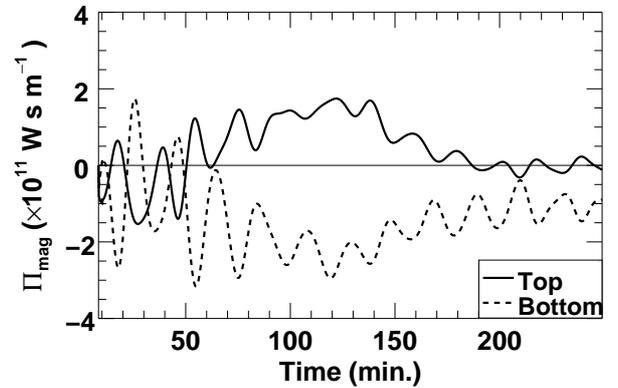}
	\caption{Time-integrated magnetic energy inflow through the boundaries of the domains indicated by orange lines in Fig. \ref{fig:setup}. The short-period component has been filtered out. \label{fig:poynting-flux}}
\end{figure}

Figure \ref{fig:time-integral} demonstrates that the increase of $W_{B}$ at the top may be related to its decrease at the bottom part of the structure. Thus, not all the energy losses at the bottom of the prominence are emitted in the form of waves. Instead, a part of this energy is transferred to the top of the structure. This transfer can explain the amplification of the oscillations at the upper part of the prominence.
We computed the incoming Poynting flux into two regions shown by the orange lines in Fig. \ref{fig:setup} as
\begin{equation}\label{poynting-flux}
\overbigdot{\Pi}_{mag}=-\oint_{\partial \Sigma} \left(\frac{B^{2}\mathbf{v}}{\mu_{0}}-\frac{(\mathbf{B\cdot v})\mathbf{B}}{\mu_{0}}\right)\mathbf{\cdot n}\ dl\ \, .
\end{equation}
Figure \ref{fig:poynting-flux} shows the time integral of the incoming Poynting flux in both regions.
We can see that during the time interval $65-150\mins$, the incoming Poynting flux at the prominence top dominates over the outcoming flux. This indicates that the magnetic energy increases at the upper prominence region. The situation is the opposite at the bottom region. The magnetic energy leaves the region, and the time integral of the Poynting flux has a negative sign. In addition, the shapes of both curves are similar but antiphase with respect to each other, indicating that the top part gains a significant part of the energy lost by the bottom part. Apart from this energy transfer, the flux at the bottom part also shows the leakage of energy emitted to the ambient atmosphere in the form of fast waves. Overall, Fig. \ref{fig:poynting-flux} indicates that magnetic energy seems to be partially transferred from the bottom to the top of the prominence. 

\begin{figure*}[!ht]
\centering\includegraphics[width=0.95\textwidth]{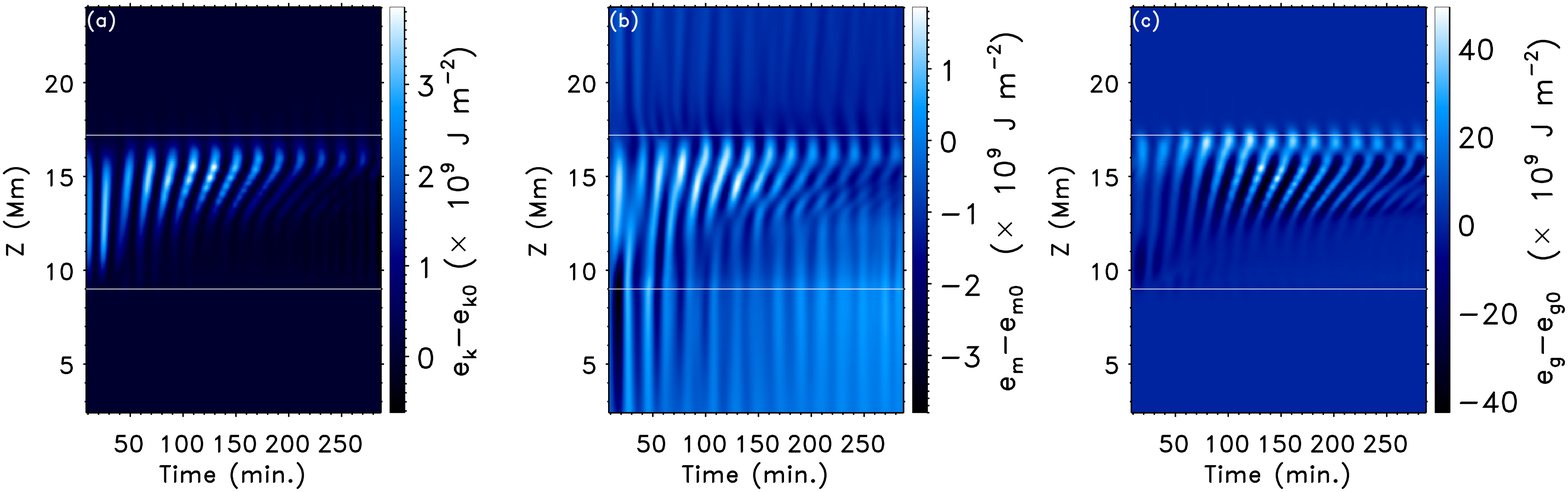}
	\caption{The time-distance diagram of the energy variations with respect to their values right before the perturbation, $e-e_{0}$, integrated in the region from $x=-12$ Mm to $x=12$ Mm. The white lines denote the heights where the prominence is located. (a) kinetic energy density, $e_{k}=\rho v^{2}/2$; (b) magnetic energy density, $e_{B}=B^{2}/2\mu_{0}$; (c) gravitational energy density, $e_{g}=\rho gz$.
	\label{fig:energies}}
\end{figure*}

 In order to understand the physical reason for the transfer of the energy from the bottom to the upper layers of the prominence, we computed the time-distance diagrams of the different contributions to the total energy along the vertical direction integrated in the region from $x=-12$ Mm to $x=12$ Mm. The result of this calculation is shown in Fig. \ref{fig:energies}. The three panels show the kinetic energy density, $e_{k}=\rho v^{2}/2$ (Fig. \ref{fig:energies}a), the magnetic energy density, $e_{B}=B^{2}/2\mu_{0}$ (Fig. \ref{fig:energies}b), and the gravitational energy density, $e_{g}=\rho gz$ (Fig. \ref{fig:energies}c). We are interested in the variation of these energies after the mass loading. Therefore, Fig. \ref{fig:energies} shows the difference with respect to the values of each of the energies at $t=8.3\mins$.
The light and dark fringes reflect the plasma motions and velocity variations. The figure clearly shows the phase shift between the oscillations at different heights inside the prominence. In addition, it can be observed that the kinetic energy increases at $t=60-130\mins$ and at heights $z>13$ Mm. A similar increase can be seen for the magnetic and gravitational energies in Figs. \ref{fig:energies}b and \ref{fig:energies}c. The magnetic energy increase is associated with the energy exchange between the bottom and top parts of the prominence, as was already shown in Fig. \ref{fig:poynting-flux}. The gravitational energy increases during the same period of time which is related to the amplification of the oscillation velocities at the top of the prominence. Since plasma is accelerated, it can reach higher heights along the magnetic field. As a result, the total gravitational energy increases. At the bottom, in the region of the strong attenuation, we can see the energy losses in all the panels. As mentioned before, some fraction of the energy is taken away by the fast MHD waves, and some other fraction is transferred to upper prominence layers.

%%%%%%%%%%%%%%%%%% END OF THE ANALYSIS OF THE ENERGY IN A FINITE VOLUME %%%%%%%%%%%%%%%%%%%%%

%%%%%%%%%%%%% BEGINNING OF THE STUDY OF CORKS %%%%%%%%%%%%%
The results above seem to indicate that the amplification of the oscillations is related to the energy transfer from the bottom to the top parts of the structure. The magnetic field structure changes with time thanks to the oscillations, and the actual trajectory of the plasma motions is not along the unperturbed magnetic field lines anymore. In a situation with a rigid magnetic field, the trajectory coincides with the field line, and the Lorentz force has no projection along the trajectory of the plasma. However, in our situation, the trajectory does not coincide with the unperturbed field lines. In such a case, the forces along the trajectory can become different from the forces projected along the magnetic field lines.
In order to understand the detailed mechanism for the amplification and damping, we study the motion of the individual fluid elements by integrating the velocity field at each time moment. We selected two plasma elements with the initial coordinates $(x,z)=(0,16.1)$ and $(0,11.8)$ Mm that correspond to the particles at the top and the central part of the prominence.
\begin{figure*}[!ht]
	\centering
	\includegraphics[width=0.95\textwidth]{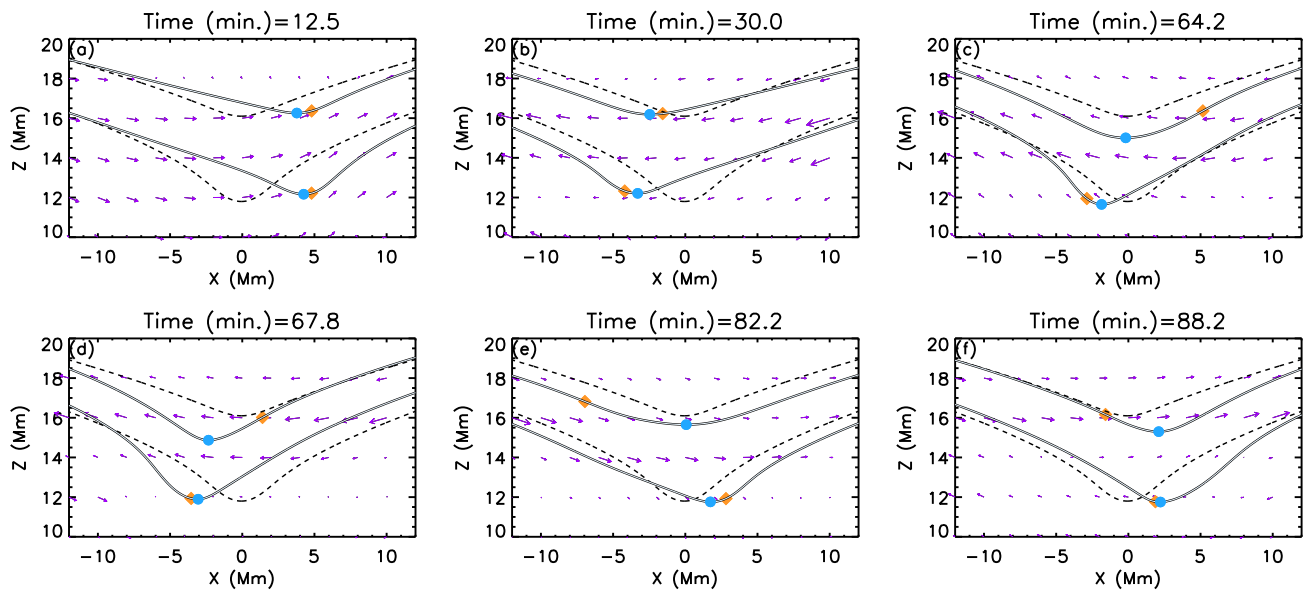}
	\caption{ Temporal evolution of the fluid elements and the magnetic field lines after the perturbation. The orange diamonds denote the positions of the fluid elements, and the blue circles correspond to the locations of the center of the dips. The purple arrows denote the velocity field. Dotted lines are unperturbed magnetic field lines; solid lines are actual magnetic field lines at a given moment. Note that the magnetic field perturbation is multiplied by a factor of $10$ for better visibility.\label{fig:evolution-corks}}
\end{figure*}
In order to highlight the modifications of the magnetic field lines due to the plasma motions, we multiply the magnetic field perturbation by a factor of $10$. This magnetic field is defined as $\mathbf{B'}=\mathbf{10(B-B_{0})+B_{0}}$, where $\mathbf{B}$ is the actual magnetic field at time $t$ and $\mathbf{B_{0}}$ is the magnetic field at $t=0\secs$. Similarly, the displacement of the fluid elements is also multiplied by the same factor in order to fulfill the frozen-in condition.
Figure \ref{fig:evolution-corks} shows the position of the fluid elements (orange diamonds) and centers of the dips (blue circles) of the magnetic field lines at the different time moments.  
Figure \ref{fig:evolution-corks}a shows the maximum displacement of the fluid elements right after the perturbation. We see that the field lines hosting the particles changes, and the position of the dips follow the motion of the particles. In Fig. \ref{fig:evolution-corks}b, the fluid elements move to the left of the structure. The dip at the lower line continues following the particle motion. In contrast, at the upper line, the dip is located ahead of the particle. In this situation, the upper particle reaches the dip in a position that is displaced to the left from the original dip. Thus, the upper particle has gained an increment in velocity during this first half period of oscillation. This process is repeated during the following period. In Fig. \ref{fig:evolution-corks}c, we observe that the particle has gained energy and that the amplitude of the oscillation is larger. In Fig. \ref{fig:evolution-corks}d, the dip has moved away from the particle again, leading to an increase of the amplitude. The same process is also produced in the reverse direction, as can be observed in Figs. \ref{fig:evolution-corks}e, f. The motion of the upper dip follows the motion of the dip at the bottom line. The reason is that the magnetic structure reacts to the bulk motion of the prominence. In this sense, the changes in the magnetic configuration at the bottom part of the structure affect the top part of the structure. 

The situation is right the opposite at the bottom of the structure. The plasma motions also modify the field lines, but the dip approaches the particle in this case. In this way, the oscillatory amplitude is reduced in each oscillation period. In Figs. \ref{fig:evolution-corks}c and \ref{fig:evolution-corks}d, we can observe the following situation: the motion of the fluid element at the bottom causes displacement of the corresponding magnetic dip. As we can see from Figs. \ref{fig:evolution-corks}e and \ref{fig:evolution-corks}f, the dip at the bottom continues following the fluid element in the next period of oscillations. Thus, the particle continues losing energy in each period, and the oscillations damp quickly.

%

%%%%%%%%%%%%%%%%%% BEGINNING OF THE ADDITIONAL SIMULATIONS %%%%%%%%%%%%%%%%%%%%%

\begin{figure}[!t]
	\centering
	\includegraphics[width=0.45\textwidth]{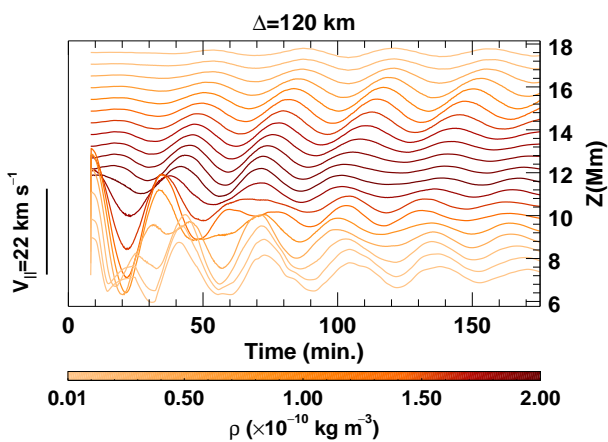}
	\caption{ Temporal evolution of $v_{\parallel}$ at the center of mass of the selected field lines in the experiment with the perturbation at the bottom. The color bar denotes the maximum initial density at each field line. The left vertical axis indicates the velocity amplitude scale. The right vertical axis denotes the height of the dips of the field lines. \label{fig:lines-perturbed-bottom}}
\end{figure}

In order to study the oscillation amplification phenomena in more detail and check how the motions at the upper and lower part of the prominence affect each other, we performed an alternative experiment described below. We used the same prominence model, but the height of the maximum of the perturbation was shifted down to $z=9$ Mm, and the characteristic vertical size of the perturbed region was reduced to $\sigma_{z}=4.8$ Mm. Using this numerical setup, we excited the prominence oscillations only at the bottom, while the upper part of the prominence remained unperturbed. Then, we repeated the calculations of $v_{\parallel}$, as it was done in Sect. \ref{sec:results}. Figure \ref{fig:lines-perturbed-bottom} shows $v_{\parallel}$ at the selected field lines. We can observe that the motions at the field lines at heights $7.2-10.2$ Mm are produced directly by the perturbation. These oscillations are significantly attenuated during the time interval shown in Fig. \ref{fig:lines-perturbed-bottom}. No perturbation has been applied for the field lines with $z>10.2$ Mm, and therefore $v_{\parallel}$ at those field lines is initially zero. However, after $20\mins$ of the evolution, we can observe a signature of oscillations with a small amplitude at the upper part. After $20\mins$, the oscillations fronts reach the top of the prominence, and their amplitude increases. At the time $100-150\mins$, we can clearly distinguish oscillation at $z>10.2$ Mm. We can also observe the phase shift of the signal between different heights that resembles Fig. \ref{fig:longitudinal-velocity}. 
We additionally performed the opposite experiment perturbing the top region and leaving the bottom and central part without initial perturbation. The analysis of motions revealed a similar result, that the oscillations at the top drives the oscillations at the bottom layers of the prominence. This experiment seems to indicate that the transfer of energy could be symmetric. However, in the regular experiments where all the layers are excited simultaneously, the bottom part of the prominence drives the motion producing the amplification of the top part. The coupling of the oscillations of the different regions of prominence is a very interesting subject for future research.
% Consequently, we conclude that pendulum motions were generated in the prominence region where they were not excited initially by the driver, thanks to the momentum transfer.
%\textbf{Additionally, we performed an experiment following the same idea but perturbing the top region and leaving the bottom and central part without initial perturbation. The analysis of motions revealed a similar result, i.e., excitation of the oscillations due to the energy transfer from the top to bottom prominence region. Consequently, we conclude that pendulum motions were generated in the prominence region where they were not excited initially by the driver, thanks to the momentum transfer.}
%This suggests that attenuation of oscillations at the bottom is partially produced by the momentum transfer and the amplification of oscillations at the top. 

In order to check if oscillation amplification is a common phenomenon of LALOs, we performed yet another experiment. In this experiment when we did not use any external perturbation but instead loaded the prominence mass at some distance from the center of the magnetic dips. Since the mass is not in equilibrium, it starts to move under the action of gravity in the direction toward the center of the dips. Thus, the plasma starts to oscillate around the equilibrium position, and the LALOs are established without any disturbance. We found that the resulting oscillations in this experiment are similar to those obtained in our regular experiment. Namely, we observe that the velocity at the top of the prominence increases while at the bottom of the prominence, it rapidly decreases. After several cycles of oscillations, the velocity of the plasma at the upper prominence region becomes high enough to allow plasma to leave the shallow dips. Consequently, the acceleration of the plasma at the top leads to mass drainage from the upper prominence region.

Summarizing all above, we conclude that the effect of amplification of the plasma at the prominence top takes place in several alternative experiments, such as experiments where we perturb only the bottom part of the prominence or do not apply perturbation but place prominence in a nonequilibrium position. This means that the effect of amplification is a frequent ingredient of LALOs and deserves more investigation in the future. 

%%%%%%%%%%%%%%%%%% END OF THE ADDITIONAL SIMULATIONS %%%%%%%%%%%%%%%%%%%%%

\section{Summary and Conclusions}\label{sec:discussion}
%%%%%%Summary of main findings
We have studied the properties of LALOs, including their periods and damping mechanism, based on 2D numerical simulations. We have used a simple magnetic field configuration that includes a dipped region. After the mass loading in the dips, we applied a perturbation to the prominence. This perturbation was directed along the magnetic field. Our main goal was to investigate the physical mechanism of the LALOs attenuation covered by the numerical diffusion. Therefore, we studied the oscillations in the same numerical model gradually increasing spatial resolution, $\Delta$. 

We have analyzed the prominence motions and computed the periods in the different prominence regions. The period of the oscillations shows an increase with height. We have also found that the period depends on $\Delta$. In the higher-resolution simulations, the period shows a strong dependence on height, while in the lower-resolution experiments, this dependence is less pronounced. The period shows a good agreement with the pendulum model in our simulations with the best resolution of 30 km.

We have studied the damping of oscillations in different experiments and at different prominence regions. In all the experiments, the prominence bottom part is characterized by strong attenuation of oscillations. This attenuation is present even in the highest-resolution experiments. The experiments with the finest resolution, $\Delta=60$ and $30$ km, demonstrated that further improvement of the spatial resolution does not significantly affect the damping time. This means that our experiments reached the resolution where the damping of oscillations is not associated anymore with the numerical dissipation but is rather caused by some physical mechanism. In a real situation, additional effects as thermal conduction and radiative losses also contribute to the damping. It is necessary to include these additional effects in order to understand which mechanism dominates in the LALOs damping. This will be the subject of future research.
%textbf{In reality, the damping time can be shorter than in our experiments due to the contribution of the other important damping mechanisms as thermal conduction and radiative losses.}

Our experiments revealed that oscillations at the prominence top are amplified during the first 130 minutes of simulations and later are slowly attenuated. The amplification appears to be more efficient and extended in time in the high-resolution experiments. 

%%%%Summary of energy study%%%%%%%%%%
In order to explain the strong attenuation of oscillations at the prominence bottom and their simultaneous amplification at the prominence top, we have analyzed the evolution of different types of energies in the corresponding regions. This analysis revealed that the damping of the oscillations is partially due to the collective work done by the gas pressure and Lorentz force. The energy is emitted in the form of fast magneto-acoustic waves. This result is in agreement with the work by \citet{Zhang:2019apj}. Furthermore, we have seen that the damping of the oscillations is related to the strength of the magnetic field. The motion of the prominence plasma produces periodic changes in the magnetic field. We have found that this effect leads to the generation of fast MHD waves. The time-distance diagram of the transverse velocity provides another piece of evidence for the wave leakage. The inclination of the wave front ridges found in the time-distance diagram is in agreement with the inclination predicted by the Alfv\'en speed. Yet another conclusion from our analysis is that the Lorentz force plays an important role in the damping and amplification of LALOs. While at the bottom, it contributes to the kinetic energy losses and acts to decelerate plasma, at the prominence top, the work done by the Lorentz force is positive and provides the gain of the energy needed for amplification of oscillations. The analysis of the Poynting flux revealed that a significant portion of the energy leaving the bottom part goes to the top. These results suggest that the energy losses at the lower prominence region are caused by both the wave leakage and the energy and momentum transfer to the upper prominence region.

%%%%%%Conclusion
Our study of LALOs based on 2D numerical simulations has shown that high spatial resolution is crucial for investigating the periods of LALOs. The period agrees with the pendulum model only when the spatial resolution is high enough. The high spatial resolution is also important for the understanding of the damping of LALOs. On the other hand, the numerical dissipation can hide the important physical mechanism as amplification of LALOs. 
%Our results provide a limit to the value of the numerical resolution required to address the physics of LALOs, which is 30 km.

%%%%%%Future prospects
In the future, it would be necessary to study the attenuation-amplification mechanisms further by using high-resolution experiments with more complex 3D setups, which would allow us to take the mechanism of resonant absorption into consideration. Furthermore, it is also desirable to include the nonadiabatic effects. This can allow us to study the relative importance of the mechanisms described in this paper with respect to the resonant absorption and nonadiabatic effects. On the other hand, more observations are needed to study further the phenomena of the amplified prominence oscillations.

\begin{acknowledgements}  V. Liakh acknowledges the support of the Instituto de Astrof\'{\i}sica de Canarias via an Astrophysicist Resident fellowship. M. Luna acknowledges support through the Ramón y Cajal fellowship RYC2018-026129-I from the Spanish Ministry of Science and Innovation, the Spanish National Research Agency (Agencia Estatal de Investigación), the European Social Fund through Operational Program FSE 2014 of Employment, Education and Training and the Universitat de les Illes Balears. V. L. and M. L. also acknowledges support from the International Space Sciences Institute (ISSI) via team 413 on “Large-Amplitude Oscillations as a Probe of Quiescent and Erupting Solar Prominences.” E. K. thanks the support by the European Research Council through the Consolidator Grant ERC-2017-CoG-771310-PI2FA and by the Spanish Ministry of Economy, Industry and Competitiveness through the
grant PGC2018-095832-B-I00 is acknowledged. V. Liakh, M. Luna, and E. Khomenko thankfully acknowledge the technical expertise and assistance provided by the Spanish Supercomputing Network (Red Espa\~{n}ola de Supercomputac\'{\i}on), as well as the computer resources used: the LaPalma Supercomputer, located at the Instituto de Astrof\'{\i}sica de Canarias. The authors thankfully acknowledge the computer resources at MareNostrum4 and the technical support provided by Barcelona Supercomputing Center (RES-AECT-2020-1-0012 and RES-AECT-2020-2-0010)
\end{acknowledgements} 
 
   \bibliographystyle{aa} % style aa.bst
   \bibliography{bibtex.bib} % your references Yourfile.bib
\end{document}